\documentstyle[amssymb,a4p,12pt,cite,epsfig]{article}
\newcommand{\udright} {\rightrightarrows}
\newcommand{\udleft} {\leftrightarrows}

\begin{document}
\renewcommand{\arraystretch}{1.6}
\begin{titlepage}
\vspace{4mm}
\bigskip

\vspace*{1.0mm}

\hspace*{11.5cm}TAUP-2791-05\\

%\centering{\large\it SEMI-FINAL DRAFT}
\vspace{7mm}

\bigskip 
\begin{center}
{\Large{\bf Measurement of low energy longitudinal polarised\\
\vspace{2mm}

positron beams via a Bhabha polarimeter}}

\vspace{17mm}

{\Large\bf\it
Gideon Alexander\footnote{Corresponding author, Physics Department,
Tel-Aviv University, Tel-Aviv, Israel. Tel.: +972-3-640-8240;
fax: +972-3-640-7932.\\
\hspace*{5mm} {\it E-mail address:} alex@lep1.tau.ac.il (G. Alexander)} 
and Erez Reinherz-Aronis
}

\vspace{6mm}

{\large
School of Physics and Astronomy\\ 
Raymond and Beverly Sackler Faculty of Exact Sciences\\
Tel-Aviv University, Tel-Aviv 69978, Israel\\
}

\vspace{10mm}  

\end{center}

\begin{abstract}
The introduction of a longitudinal polarised positron beam 
in an $e^+e^-$  linear collider calls for
its polarisation monitoring and measurement at low energies near its
production location. Here it is shown that a relatively simple 
Bhabha scattering
polarimeter allows, at energies below 5000 MeV, a more than adequate
positron beam longitudinal polarisation measurement by
using only the final state electrons. 
It is further shown that 
%In particular the question which
out of the three, 10, 250 or 5000 MeV positron beam 
energy locations, where the polarisation
measurement in the TESLA linear
collider can be performed, the 250 MeV site is best suited for this task.  
\end{abstract}
\vspace{1.0cm}
\begin{center}
(\today)
\end{center}
\vspace{5mm}

\begin{flushleft}
\vspace{2mm}

\end{flushleft}
\end{titlepage}
\section{Introduction}
Over more than two decades plans to construct high energy
$e^+e^-$ linear colliders, reaching the centre of
mass (CM) energy up to around 1 TeV, have been studied in many
institutes of high learnings and research laboratories. Whereas
in the beginning several designs for the collider have been
pursued in parallel (see e.g. Refs. \cite{tesla,glc,nlc}),
recently it has been world wide agreed upon that only one International
Linear Collider (ILC) should be planned and constructed.  
The main motivation to build such a collider is to further 
our knowledge on the physics of particles and fields, 
and in particular to explore the still missing Higgs sector of the
Standard Model and to search for phenomena beyond this model
like the existence of super-symmetric particles.\\ 

In assessing the research and
discovery power of such a collider it has soon been realised that
one should try and equip it, not only 
with a longitudinal polarised electron 
beam, but also with a longitudinal polarised positron beam \cite{case}. 
To achieve 
this goal one has to develop methods to polarise longitudinally
the electron and positron beams and to have the needed devices to
measure and maintain their  polarisation levels. As for the electron beam,
the polarisation production can be achieved e.g. by irradiating GaAs crystals
with circular polarised laser beam so as to emit low energy 
longitudinal polarised
electrons which are further linearly accelerated to the desired
energy.
This method has already been successfully applied to the SLAC linear collider,
the SLC, which operated with a $\sim$73$\%$ longitudinal polarised 
electron beam at the laboratory energy of 45.6 GeV \cite{ahe}.
%Several methods to produce a 
%longitudinal polarised positron beam are still
%under study. 
One attractive proposition for the production of a
longitudinal
polarised positron beam  
is the undulator based method 
\cite{undulator}
which is currently tested
by the E166 experiment \cite{slac} at SLAC. 
In that experiment a $\sim$50 GeV
electron beam
passes through a helical undulator to produce circular polarised photons
which create in a target $e^+e^-$ pairs. 
These pairs divide between themselves  
the polarisation of the photons in proportion to their 
relative momentum.
Thus, by selecting the more energetic outgoing positrons one should be able to
build up a longitudinal polarised beam for linear colliders.\\

Polarisation measurement of a positron beam, at or near to its creation
position, is needed for the routine operation
of the collider, firstly to verify the presence of polarisation and secondly
to guide the accelerator operators 
in their efforts to maximise the polarisation level.
At the same time the precision required for this measurement does not
have to be  
as high as 0.5$\%$ which is expected to be reached 
via a Compton polarimeter \cite{schuler}
at the $e^+e^-$ interaction point where the data for 
particle physics evaluation is collected. 
Therefore the proposition
to install at, or near, the production of the polarised positron beam
a fixed magnetised iron target polarimeter may well be 
an attractive proposition.  
In the present work we investigate the feasibility to 
measure the longitudinal polarisation
of a positron beam at or near to its creation via a fixed magnetised iron target
which we here will refer to as a Bhabha polarimeter.

\section{The physics background}

\begin{figure}[ht]
\centering{\epsfig{file=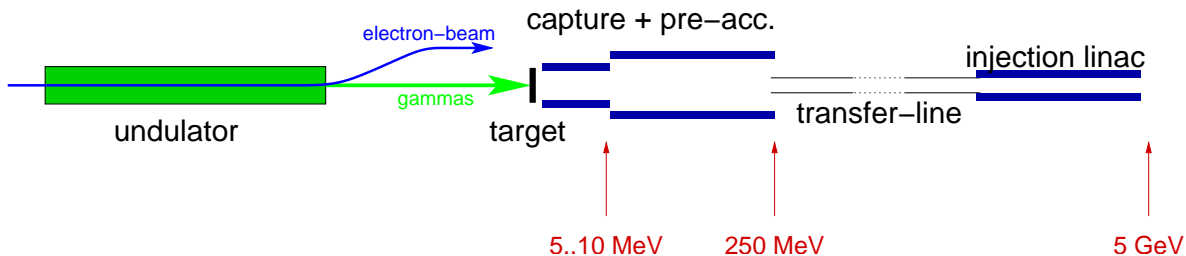,height=3.3cm}} 
\caption{\small A schematic layout of the production of a longitudinal
polarised positron beam in a linear collider as outlined in 
Ref. \cite{tesla}. 
Marked by the arrows are the locations and their
corresponding beam energies where
a longitudinal polarimeter may be installed.}
\label{injector}
\end{figure}

The energy region which we here consider for the
positron beam polarisation measurements, near its production
in a linear collider, follows the design
put forward by the TESLA project \cite{tesla} which  
schematically is shown in Fig. \ref{injector}.
In this plan the positron beam is produced at energies between 5 to 10
MeV by the circular polarised photons emerging from an electron beam 
passing through a helical
undulator. In the initial acceleration stages two more positions are
indicated as possible locations for a  polarimeter installation, 
one at 250 MeV and the other at 5000 MeV.\\ 
 
For the present study we select these three energy values i.e., 10, 250
and 5000 MeV, to represent those  
which eventually 
will be fixed by the final ILC design. To note is that 
even at 5000 MeV laboratory energy, the centre of mass energy of
a positron 
hitting an electron in a fixed iron target of 
a Bhabha polarimeter is only $\sim$71.4 MeV. That is,
far below the mass energy of the Z$^0$ gauge boson. 
Thus it is here sufficient to describe
the Bhabha scattering process of longitudinal polarised positrons and
electrons in terms of the QED diagrams involving photon 
exchanges.
The Bhabha process with longitudinal and transverse polarised
beams is e.g. formulated in Ref. \cite{renard} for the case where only the QED
processes contribute. In  Ref. \cite{riemann} the Bhabha
scattering of longitudinal polarised beams is dealt with
in terms of a semi-analytical approach realised by a Fortran program 
{\it Zfitter} which is valid at least
up to $\sqrt{s}=350$ GeV, at or 
near the threshold energy of the  $t \bar t$ quark$-$pair production. 
In this approach
all the QED as well as the Z$^0$ exchange processes and their interferences
are included.\\ 

For the CM differential Bhabha scattering cross section   
at low energies, where the Z$^0$ contribution
and the running nature of the QED coupling constant can be ignored,
we deduce from Ref. \cite{riemann}, in the absence of 
a transverse polarisation component, the following expression
\begin{equation}
%\begin{split}
\frac{d\sigma}{d\cos\theta}\ =\ \frac{\pi\alpha^2}{2s}
%\biggl\{
\left[ \lambda_1(1+\cos^2\theta)
+2\lambda_1\frac{(1+\cos\theta)^2}{(1-\cos\theta)^2}+8\lambda_3\frac{1}{(1-\cos\theta)^2}
-2\lambda_1\frac{(1+\cos\theta)^2}{(1-\cos\theta)}\right]\ .
%\biggr\}\ .
%\end{split}
\label{bhapol}
\end{equation}
Here $\theta$ is the polar centre of mass scattering angle of the
positron and
$$\lambda_1\ =\ 1\ -\ \vec{P}_+\vec{ P}_-\ \ \ \  {\rm{and}}\ \ \ \ 
\lambda_3\ =\ 1\ +\ \vec{P}_+\vec{P}_-\ ,$$
where $\vec{P}_+$ and $\vec{P}_-$ are respectively the  
longitudinal
polarisation vector of the colliding positron beam and the electrons in the
target which are defined 
in the range $0 \leq |\vec{P}_i| \leq 1$. 
If one or both of the electron and positron beams are unpolarised,
then one has $\lambda_1=\lambda_3=1$,
\begin{figure}[ht] 
\centering{\epsfig{file=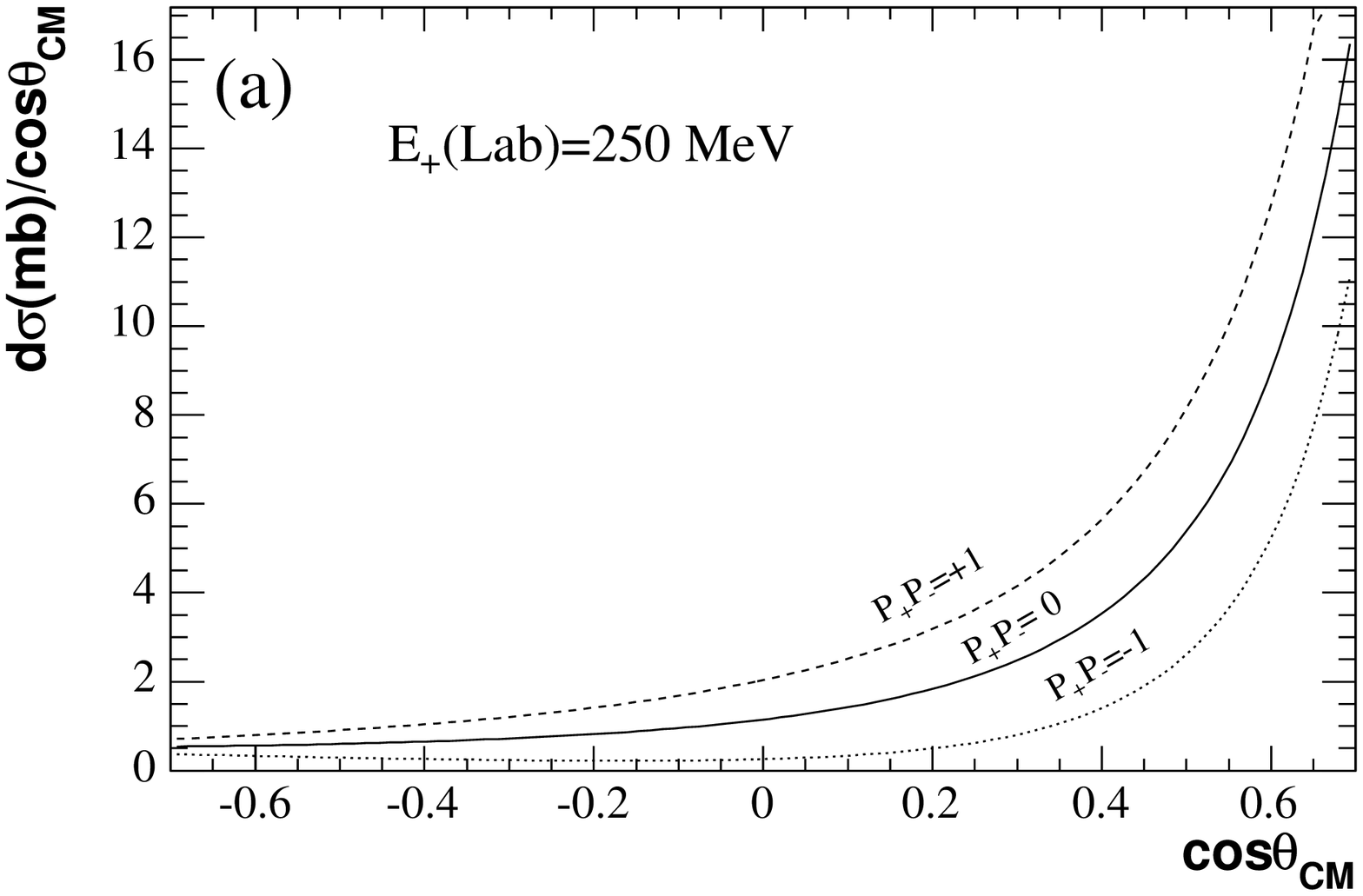,height=5.5cm}\ 
\epsfig{file=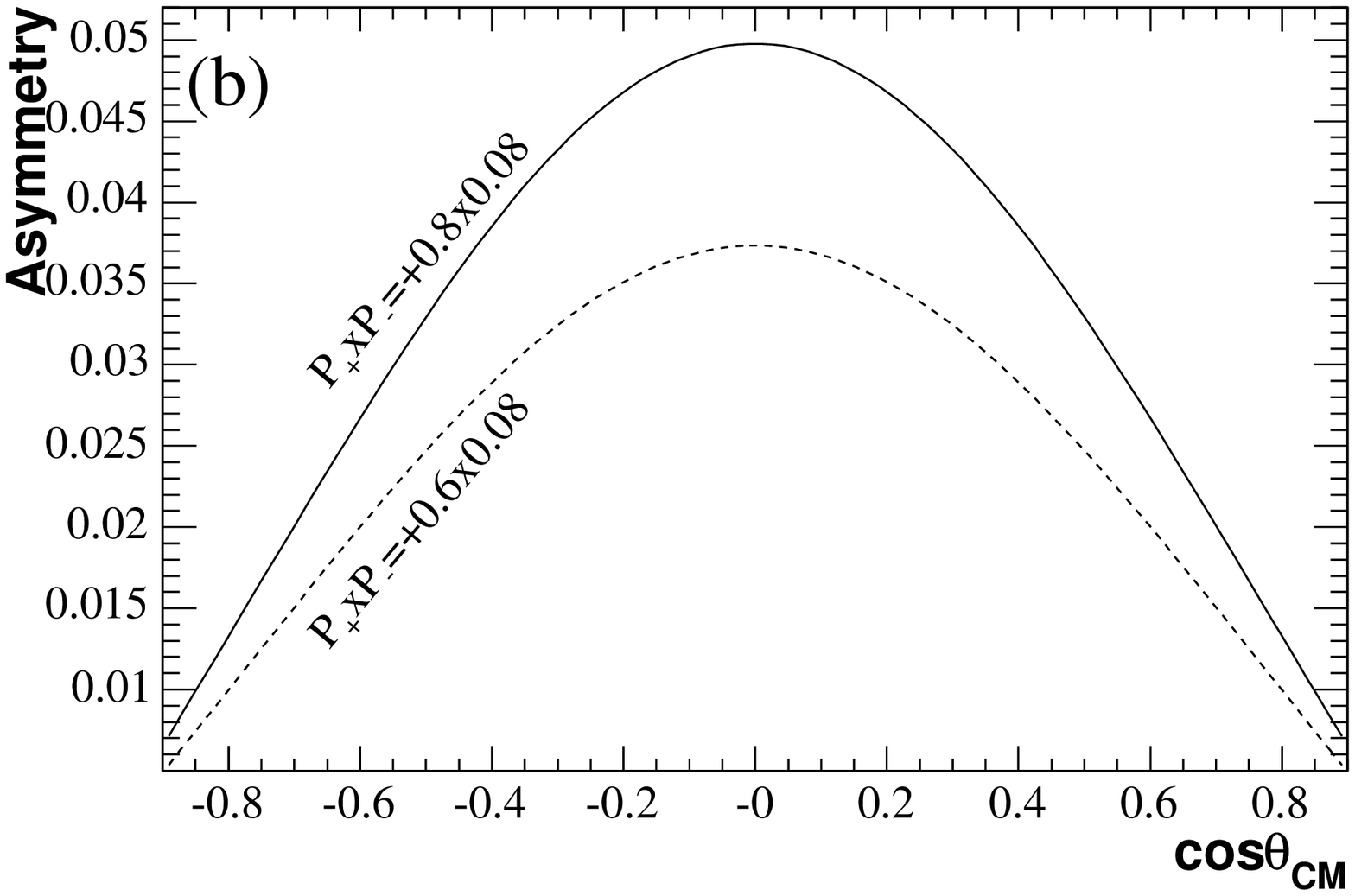,height=5.5cm}}
\caption{\small 
(a) The differential cross sections 
of the Bhabha 
scattering at 250 MeV beam energy as a function of $\cos\theta_{CM}$ 
for three longitudinal polarisation values which display the analysing
power of a Bhabha polarimeter.
(b)
The Bhabha scattering asymmetry $A$ versus $\cos\theta_{CM}$. 
The continuous and dashed lines represent the polarisation 
values of $\vec{P}_+\vec{P}_- = +0.8\times$0.08 and +0.6$\times$0.08 
respectively. 
}
\label{fig_riem} 
\end{figure}
so that the Bhabha differential cross section is reduced to
\begin{equation}
%\begin{split}
\frac{d\sigma(P_+P_-=0)}{d\cos\theta}=\frac{\pi\alpha^2}{2s}
\frac{(3+\cos^2\theta)^2}{(1-\cos\theta)^2} .
%\end{split}
\label{nopol}
\end{equation}
To illustrate the polarisation analysing power 
given by Eq. (\ref{bhapol}) we show in 
Fig. \ref{fig_riem}a the Bhabha scattering differential cross
section at 250 MeV beam energy for two limits of the 
longitudinal polarisation states
i.e., for $\vec{P}_+\vec{P}_-$ equal to +1 and
$-$1 and for the zero polarisation case. As can be seen, there is a substantial difference
between the magnitudes of the differential Bhabha scattering
for these three polarisation states which
can be utilised for the polarisation measurement.
In practice however,
the iron polarisation level cannot exceed the value of 0.08 and
it is judged that the positron beam polarisation will 
not reach levels higher than $\sim$0.6.\\

\begin{figure}[h]
\centering{\epsfig{file=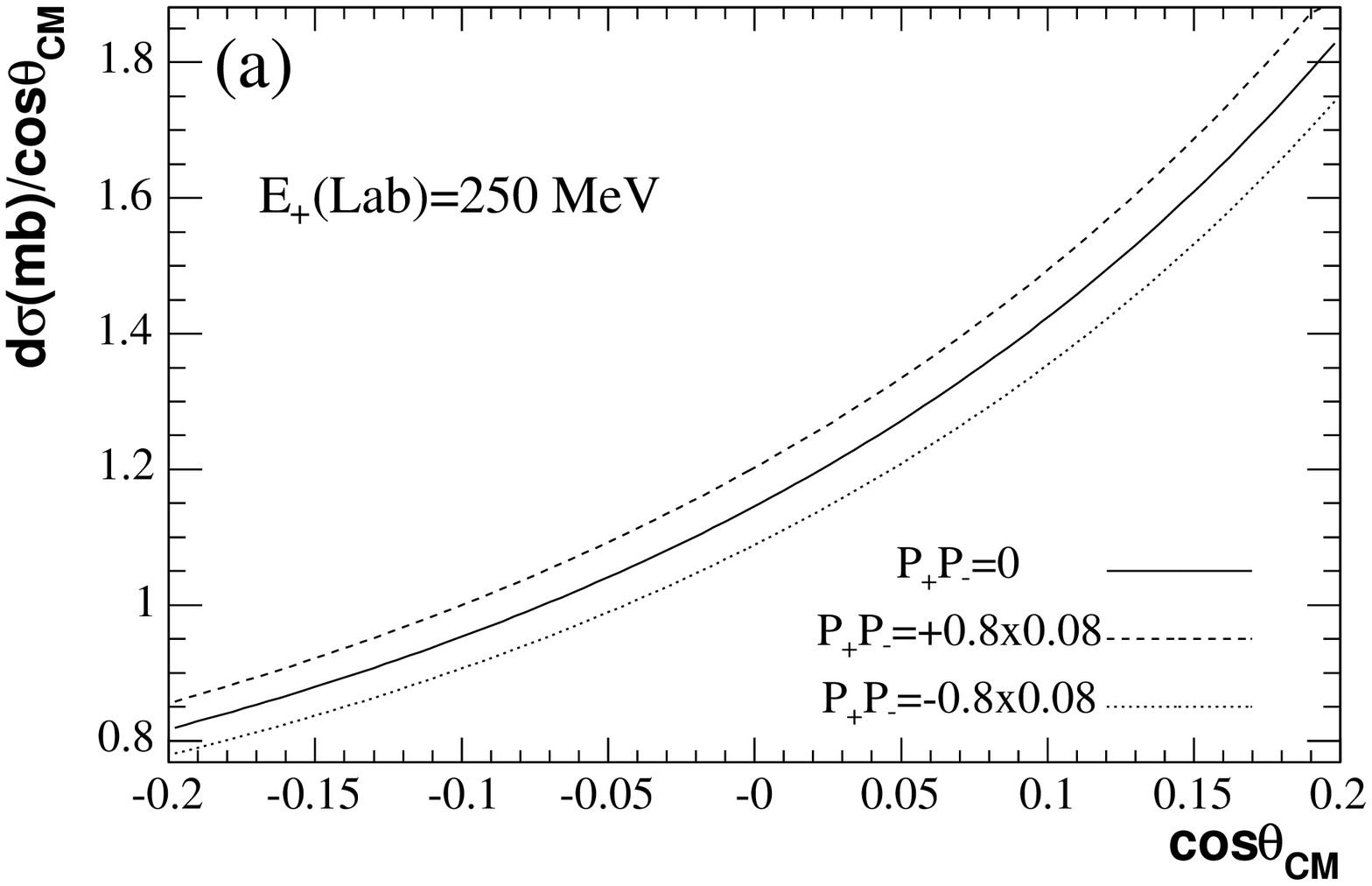,height=5.5cm}\ \ \ \
\centering{\epsfig{file=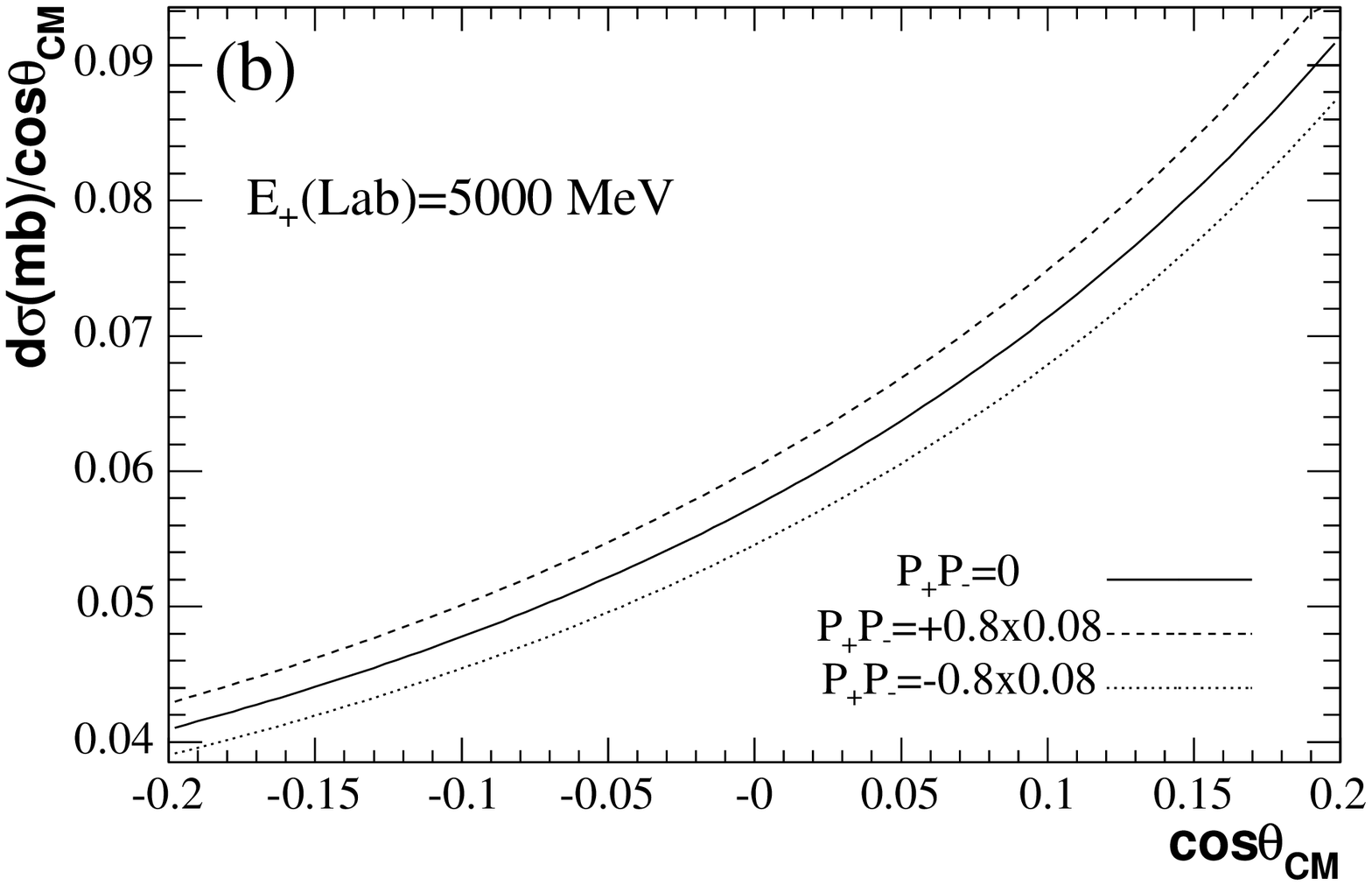,height=5.5cm}}}
\caption{\small The differential Bhabha cross section in the centre of
mass system
as a function of $\cos\theta_{CM}$ at laboratory positron beam energy
of (a) 250 and (b) 5000 MeV having three polarisation levels
of $\pm$0.8 and zero while the iron target polarisation level is
0.08.} 
\label{fig_diff}
\end{figure}

The measurable asymmetry $A$, which is a
function of $x~(\equiv \cos\theta_{CM})$, is given in the centre of mass
system by
\begin{equation}
A\ =\ \frac{d\sigma/dx(\udright)-d\sigma/dx(\udleft)}
{d\sigma/dx(\udright)+d\sigma/dx(\udleft)}
\ =\ \frac{7-6x^2-x^4}{(3+x^2)^2}\vec{P}_+\vec{P}_- \   \ 
\stackrel{\theta =\pi/2}{\overrightarrow{\hspace{1.5cm}}}\ 
\frac{7}{9}\vec{P}_+\vec{P}_- 
\label{maple}
\end{equation}
which is shown, as a function of
$\cos\theta_{CM}$, in Fig. \ref{fig_riem}b for the 
combined polarisation values
of $\vec{P}_+\vec{P}_- = +0.8\times 0.08$ and $+0.6\times0.08$.
Here we denote by $(\udright)$ and $(\udleft)$ respectively
the states where the positron beam polarisation is parallel 
and anti-parallel
to the target electron polarisation direction.  
This asymmetry behaviour is identical to that one found for the
M\o ller scattering (see e.g. Ref. \cite{alex}) where
at $\cos\theta_{CM}=0$ reaches the maximum value of 
$(7/9)\times \vec{P}_+\vec{P}_-$.
Assuming the beam polarisation 
to reach the high value of 0.8, then the resulting differential cross
section are shown in Fig. \ref{fig_diff} at 250 and 5000 MeV beam energy
for three polarisation states.

\subsection{The features of a polarimeter setup}
Two of the main elements of a Bhabha (or M\o ller) polarimeter consist of 
a magnetised iron, or iron alloy, target and a detector system for
the identification and recording of the scattering process.
In the present work 
we consider an iron target of a 10$\mu$m, a width  which has   
been previously successfully applied to a M\o ller polarimeter \cite{moller}, 
in order to reduce as much as possible secondary interactions
and other background sources like the bremsstrahlung. 
The target, which is
cooled down to $\sim$110 K, is placed in a magnetic field to
reach the polarisation level of about 8$\%$ which is
its maximum possible value due to the 
fact that only two out of the iron 26 electrons can be polarised.
In a high magnetic field, of about four Tesla, it was shown that
a target made out of thin iron foils, can be polarised {\it out-of-plane}
in saturation \cite{moller,puchi} i.e., parallel to the charged lepton
beam direction. 
At moderate and low
magnetic fields the polarisation direction is found to lie in the plane
of the target face. In this case the target has to be tilted in the
direction of the beam in order to increase the $|\vec{P}_+\vec{P}_-|$ value
to achieve a sufficiently high polarisation analysing power. This tilt however  
increases the actual target thickness for example, from 10$\mu$m 
to 29.2$\mu$m,  
when the target is placed at 20$^0$ with respect to the beam
direction. Since the introduction 
of a high magnetic field of the order of $\sim$4 Tesla into the
accelerating domain 
may be prohibited, we have
taken for our Monte Carlo simulation study the 
less favourable target thickness of 30$\mu$m.\\
 
Here it should be noted 
that during the operation of a Bhabha polarimeter
the iron target should not heat up and with it, reduce or 
completely loose, its
polarisation. This heating problem, which depends among others factors
on the beam current and its structure and on the measurement
duration, can be kept under control even at relatively high currents of
several tens of $\mu$A (see e.g. Ref. \cite{alex}). In any case, 
an online measurement of the relative iron foil polarisation during
the polarimeter operation should be carried out for example 
with a laser beam making use of the polar Kerr effect \cite{puchi}.\\ 

Further we envisage that the Bhabha scattering outgoing charged leptons
are steered into the polarimeter detector via a magnetic field which
allows one to separate the electrons from the positrons 
and prevents the outgoing photons from hitting the detector.
For the recording of the
Bhabha events we foresee a pixel detector 
which covers a sizable part of the azimuthal angle region and an adequate polar
angle range in the laboratory angular region which we here,
in our feasibility study,
set to be the one corresponding to
the centre of mass cosine angle of $-$0.65 to +0.4. The
setting of these limits at the corresponding laboratory angles
can be realised for example by adjustable collimators similar
to the ones applied previously to 
a M\o ller polarimeter \cite{moller} which 
selected the range of scattering angles and did cut off
electrons at both smaller and larger angles. Furthermore, in 
that polarimeter setup, in front of
the detector two slits were placed to define the actual 
acceptance of the polarimeter. 
However,
as will be shown later, unlike the case of a M\o ller polarimeter,
the need for an energy measurement of the outgoing
electrons may be relaxed in
a Bhabha polarimeter.
The dimension of
Bhabha polarimeter detector will have eventually to be determined 
by its distance   
from the target and the angular spread 
caused by the specific magnetic-optic system to be used. Finally the 
number of pixels and their size,  
is above all dictated by the need to keep
the multiple-hit pixels to an
insignificant number between the readout times of the detector.\\ 

\section{Monte Carlo simulation} 
The features of  Bhabha polarimeter using a fixed iron
target of 30$\mu$m width operating with positron beams having the 
energies of 
10, 250 and 5000 MeV 
were simulated via the GEANT4 Monte Carlo program 
\cite{geant4,geant_man} which is currently void of spin effects. 
As a consequence our study on the Bhabha polarimeter sensitivity
to the angular distribution of the scattering cross section and its 
strength, is
carried out in the vicinity of zero polarisation. However judging from
Fig. \ref{fig_riem}a where the Bhabha scattering dependence on 
$\cos\theta_{CM}$ at 250 MeV is shown for the two extreme polarisation
cases of $P_+P_- = \pm 1$ and remembering that in practice
$P_- = 0.08$, the features of the polarimeter
should essentially be independent of the positron polarisation level.      
%This shortcoming of the Monte Carlo program meant that our studies
%on the sensitivity of the Bhabha polarimeter to angular distributions
%and scattering cross section strength in the vicinity of zero
%polarisation.    
In the simulations  
200 Million positrons did hit the polarimeter target
in each of the above
selected beam energies 
and the outgoing
positive and negative particles were recorded.  
\begin{figure}[ht]
\centering{\epsfig{file=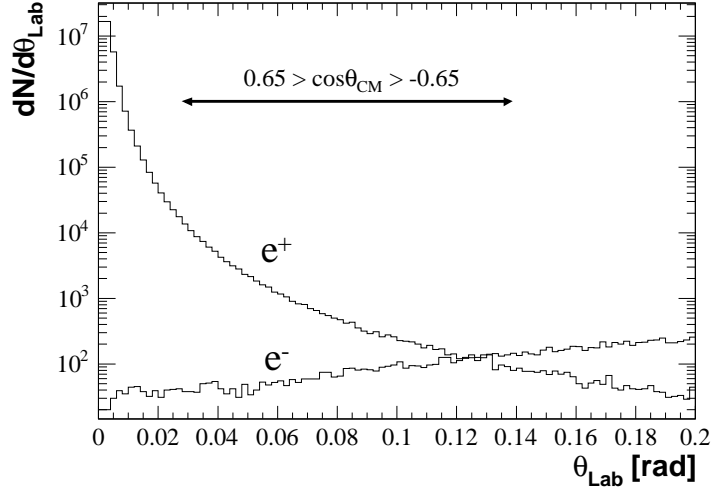,height=7cm}} 
\caption{\small 
The laboratory $\theta_{Lab}$
angular distributions, in the range 0 to 0.2 rad, 
of the outgoing electrons and positrons 
as obtained from a GEANT4 Monte Carlo generated sample of a 250 MeV 
positron beam hitting an iron target of a 30$\mu$m thickness.
The horizontal solid line marks the range corresponding to 
the CM cosine angle between $-$0.65 and +0.65. 
}
\label{labe+e-}
\end{figure}
It was found out that at 10 and 250 MeV essentially all
the outgoing charged particles were electron and positrons. 
At the beam energy of 5000 MeV some
charged final states hadrons were also observed. 
In Fig. \ref{labe+e-} the laboratory $\theta_{Lab}$ angular distribution
at 250 MeV beam energy is shown  
for the outgoing positrons and electrons.
Within the range of $\theta_{Lab}$ of 0.1377 and 0.0293 rad, 
which corresponds in the CM to the region 
$-0.65 \leq \cos\theta_{CM} \leq +0.65$,
one observes that the ratio of positrons to electrons is 
about 19. This ratio and those found at 10 and 5000 MeV beam energies  
are listed in Table \ref{table1} where they are seen to lie in
the range of $\sim$20 to $\sim$16. From the Monte Carlo studies
follows that these large ratios, which in the absence of background
should be equal to one, are 
mainly due to the
contribution of the bremsstrahlung process which contains in its
final state a positron. To eliminate this dominant background
we further
restrict our analysis to  
the detected outgoing electrons and show that these are
sufficient to identify the Bhabha scattering process and 
to measure the beam polarisation.\\ 
\renewcommand{\arraystretch}{1.1}
\begin{table}[h]
\caption{\small Monte Carlo results for the ratio of
the outgoing positrons to the outgoing electrons, and the ratio between
the background (BG) electrons to the total outgoing electrons 
emerging within the centre of mass symmetric polar angle
range of $-0.65 \leq \cos\theta_{CM} \leq +0.65$. 
The data are shown
for the three positron beam laboratory energies of 
10, 250 and 5000 MeV 
impinging on an iron target of 30$\mu$m effective width. The values 
given in the brackets
refers to a target width of 10$\mu$m 
thickness. 
}
\begin{center}
\begin{tabular}{||c||c|c|c||}
\hline\hline
Beam energy [MeV]& $\theta_{Lab}$ range [rad] &
No. e$^+$/No. e$^-$&Fraction of BG e$^-$\cr 
\hline
10 & $0.5940\ -\ 0.1424$ &19.9 & $--$\cr
250& $0.1377\ -\ 0.0293$ &$$18.9 & 6.3$\%$\ \ (2.0$\%$)\cr
5000 &$0.0310\ -\ 0.0066$ &16.5 & 64$\%$\ \ (22.5$)\%$\cr
\hline\hline
\end{tabular}
\end{center}
\label{table1}
\end{table}
\renewcommand{\arraystretch}{1.1}

\begin{figure}[t]
\centering{\epsfig{file=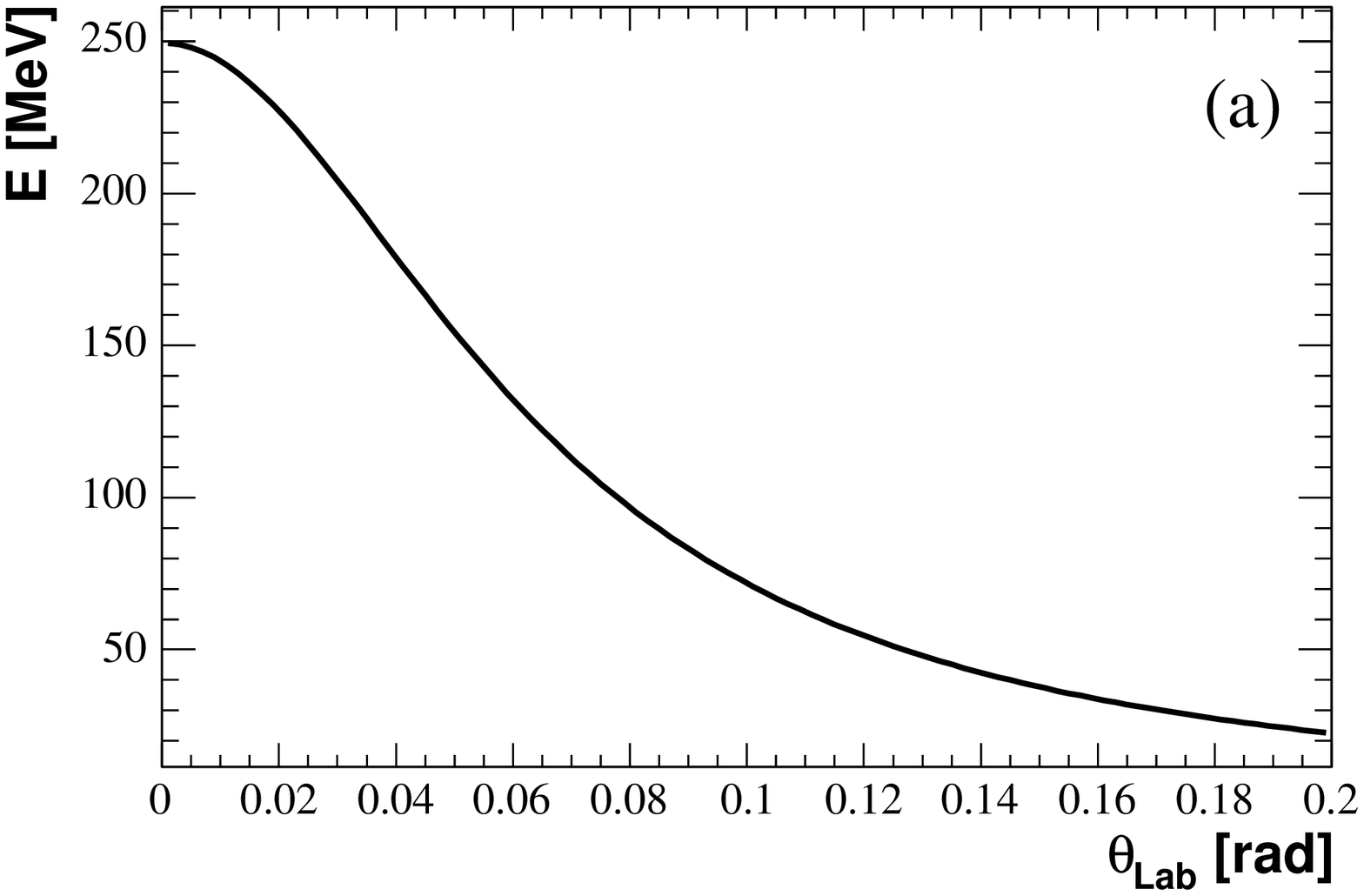,height=5.4cm}
\epsfig{file=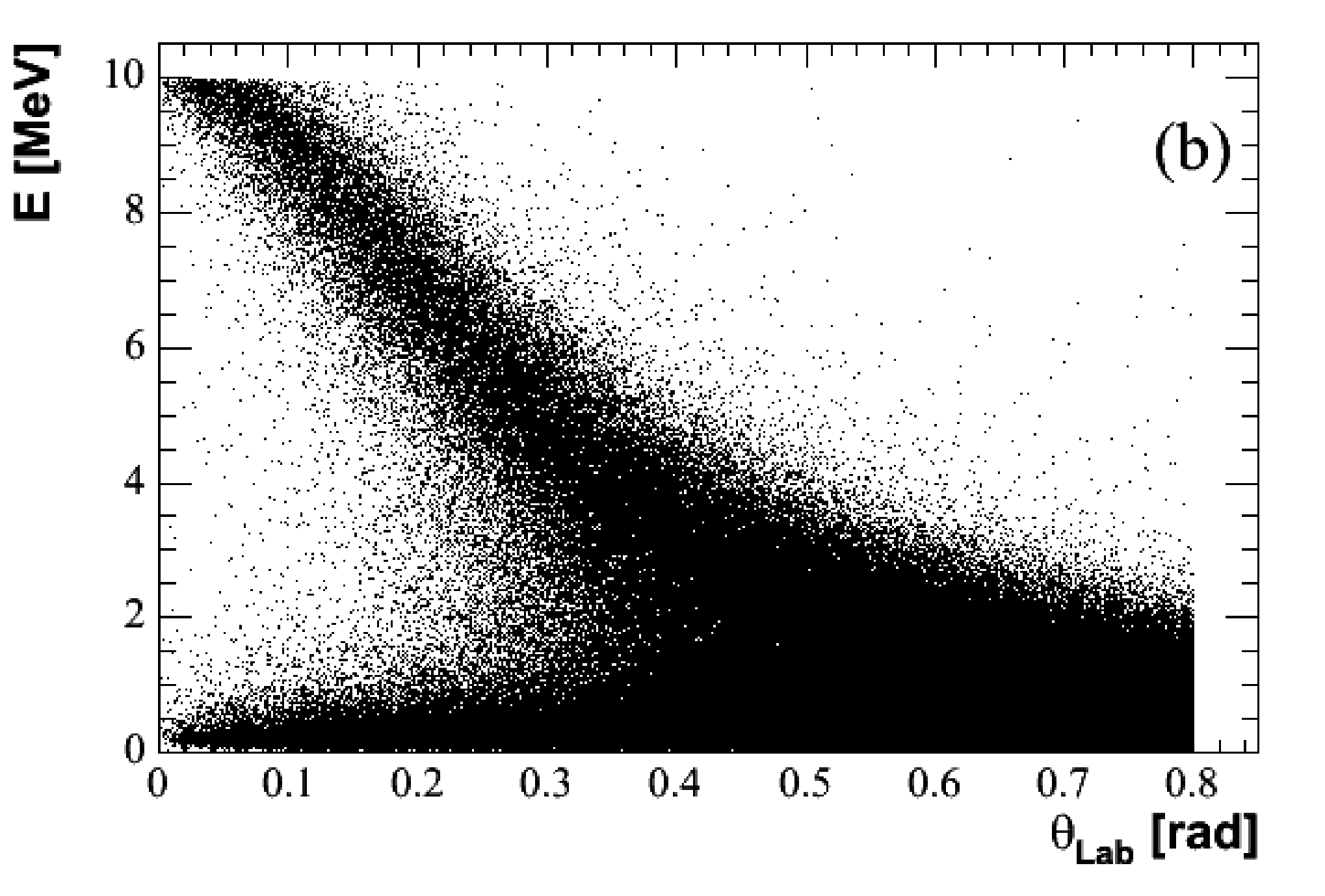,height=5.4cm,width=8.0cm,
bbllx=2pt,bblly=2pt,bburx=585pt,bbury=394,clip=}\\
\epsfig{file=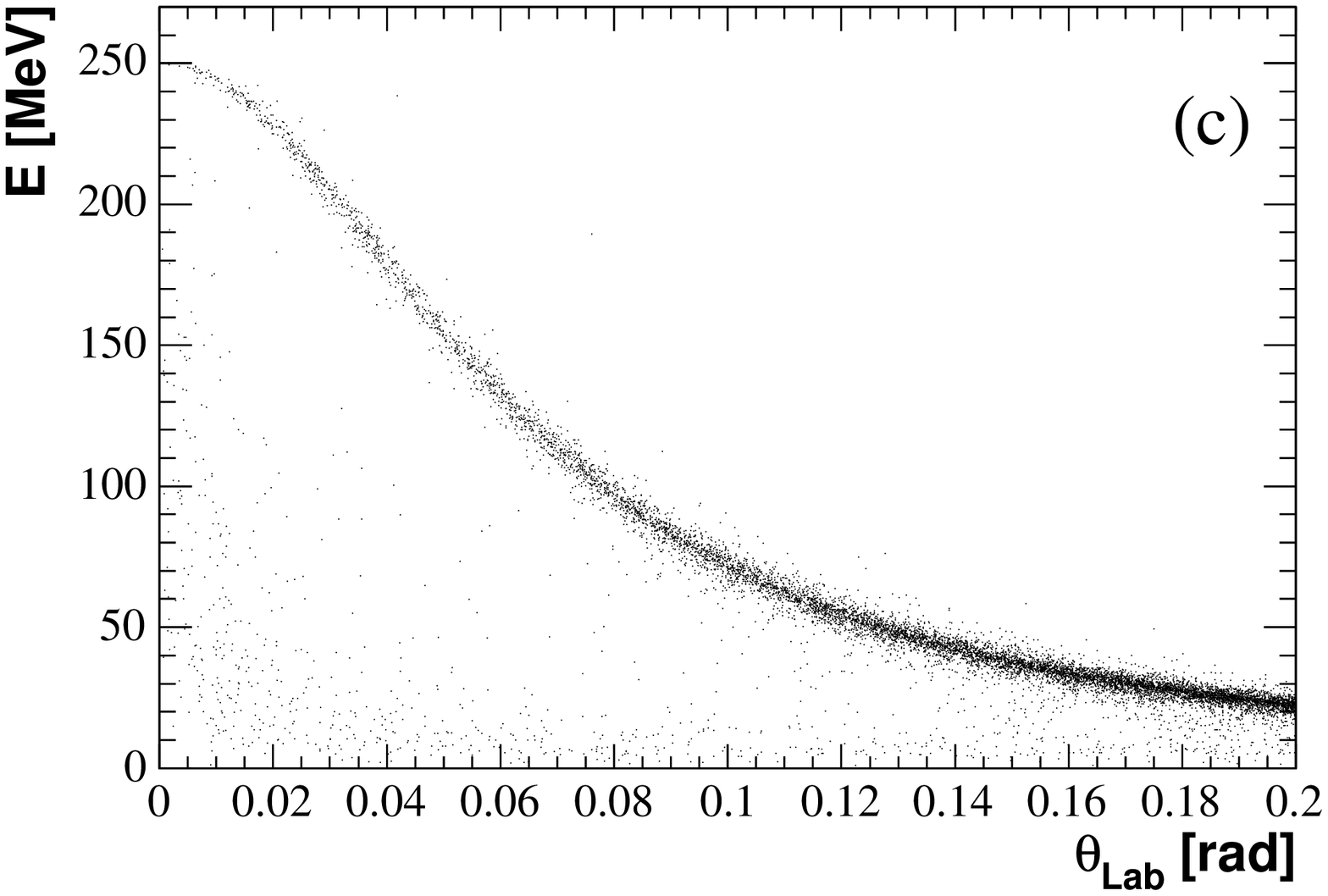,height=5.4cm}
\epsfig{file=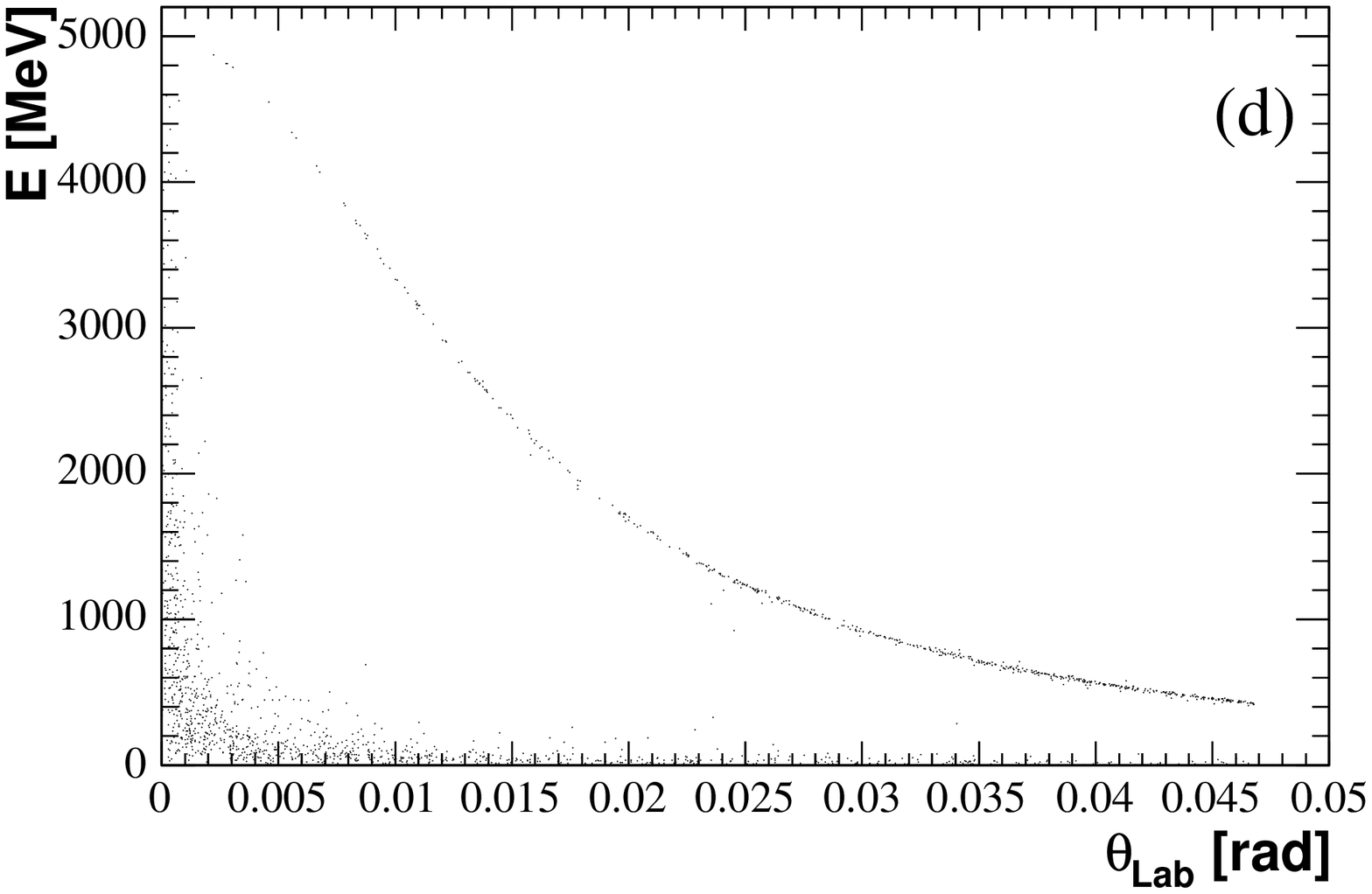,height=5.4cm}} 
\caption{\small 
The Bhabha polarimeter response to the outgoing electrons 
laboratory energy 
on their scattering angle
$\theta_{Lab}$ in the range corresponding to $\theta_{CM}=0$ to 2.55 rad.
(a) The dependence as calculated from Eq. (\ref{evstheta}), 
at beam energy of 250 MeV. The scatter plots (b), (c) and (d) were
obtained from GEANT4  
Monte Carlo generated electrons samples produced at each energy by 
200 Million positrons hitting a
30$\mu$m thick fixed iron target. The plots (b), (c), and (d)
correspond respectively to  
10, 250 and 5000 MeV beam energy.
}
\label{ek}
\end{figure}

Next we turn to the relation between the laboratory energy of the 
outgoing electrons and their scattering
angle $\theta_{Lab}$ which is given, in terms of $\cos\theta_{Lab}$, by
\begin{equation}
E_{Lab}\ =\ m_e\frac{\gamma+1 +(\gamma-1)\cos^2\theta_{Lab}}
{\gamma+1 -(\gamma-1)\cos^2\theta_{Lab}}\ .
\label{evstheta}
\end{equation}
This relation is shown in Fig. \ref{ek}a  
for a positron beam energy of 250 MeV. 
In Figs. \ref{ek}b, \ref{ek}c and \ref{ek}d are shown the 
Monte Carlo generated 
scatter plots of the laboratory energy of the negative outgoing
leptons versus 
their angle $\theta_{Lab}$ for respectively 10, 250 and 5000 MeV
incident positron beam energy. Each scatter plot was produced
by $200\times$$10^6$ positrons impinging on an iron target.\\
 
By comparing at 250 MeV the Monte Carlo  results with the calculated
distribution one univocally identifies  
the dense band, which is well separated from the background, 
as belonging to the
Bhabha scattering process. This situation is also true for the
5000 MeV beam energy scatter plot shown in
Fig. \ref{ek}d where the background is lying even further away from the
Bhabha signal. 
At 10 MeV beam energy however the isolation of the Bhabha scattering signal 
is severely hampered by the large background which is seen to merge
with the signal at about $\theta_{Lab} \gtrsim 0.25$ rad. From the Monte
Carlo studies the low energy background, seen in all the three energy scatter
plots, stems mainly from Compton scattering and $e^+e^-$ pairs produced 
in the iron target by soft
secondary photons the amount of which  
is seen to be approximately the same at 250 and 5000 MeV
incoming beam energy. 
The higher energy background seen at small
angles is coming from 
Bhabha scattering events where the outgoing electron
suffered further on energy loss before emerging out of the target.  
As expected, the Bhabha scattering signal is
smaller at 5000 MeV 
than at 250 MeV and as a consequence the
signal to background ratio is also smaller as the beam energy increases.
In view of all these observations it is safe to conclude that
the option to place a Bhabha polarimeter in the region where the beam
energy is 10 MeV or less is clearly disfavoured and should, if possible, 
be avoided.
Both the 250 and 5000 MeV locations, which are a priori
suitable for the installation of a Bhabha polarimeter,
are further explored in the next subsection.

\subsection{The Bhabha scattering signals and their background} 
As has been shown in Fig. \ref{ek}c and \ref{ek}d, the 
Bhabha scattering events
are concentrated
along a band lying in the laboratory energy versus $\theta_{Lab}$
plane which are well separated  from a background   
which is seen to be mainly concentrated at low 
energies and at the $(E_{Lab},\theta_{Lab})=(0,0)$ corner. 
In the $\theta_{Lab}$ region chosen here for the polarisation analyses,
which corresponds to the range 
$-0.65\leq \cos\theta_{CM} \leq +0.4$, 
this background amounts only to a about
5.2$\%$ in the 250 MeV case and to $\sim$57$\%$ in the 5000 MeV
case, out of the total number of the outgoing electrons. Therefore 
if one insists on the 5000 MeV
location for the polarisation measuements the 
amount of background should be reduced  
e.g. by introducing  
an appropriate combined energy$-$angle cut.
This option will require however  
a more elaborated experimental setup, like the one 
chosen for the M\o ller polarimeter
described in Ref. \cite{moller},
which allows to measure simultaneously the individual outgoing electrons
angle and energy and not just simply count
them within a predetermined angular sector. Here it is worthwhile to
note the the reduction of the target width to 10$\mu$m is still
leaving the background at the relative high level of 22.5$\%$.\\
   
\begin{figure}[ht]
\centering{\epsfig{file=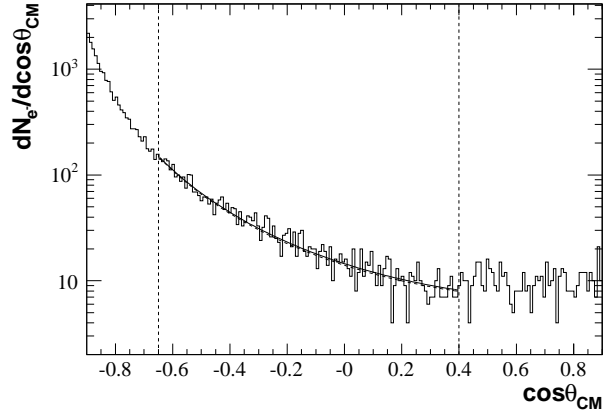,height=5.9cm}}
\caption{\small
The centre of mass angular distribution of the outgoing electrons 
obtained from a GEANT4
Monte Carlo generated event sample of a 250 MeV
positron beam hitting a
30$\mu$m thick fixed iron target. 
The solid line represents the fit results of Eq. (\ref{bhapol}) 
to the data in the $\cos\theta_{CM}$ range marked by the vertical 
dotted lines.
}
\label{coscm}
\end{figure}

To reaffirm the origin of the events in the bands shown in Fig. \ref{ek}c  
we proceeded to analyse the angular distribution of the outgoing
negative particles seen at 250 MeV beam energy in the 
centre of mass system assuming all of them to be electrons emerging
from an unperturbed Bhabha scattering process. In this case  
the transformation from the Lab angle to the CM angle is given by
\begin{equation} 
\cos\theta_{CM}= \frac{2-(\gamma+1)\tan^2\theta_{Lab}}
{2+(\gamma+1)\tan^2\theta_{Lab}}
\end{equation}
where $\gamma = E_{Lab}^{beam}/m_e$. 
The angular distribution of the
the Monte Carlo generated outgoing electrons in the CM system 
is shown in Fig. \ref{coscm}.
The solid line in the figure represents the results of a fit 
of Eq. (\ref{bhapol}) to the Monte Carlo generated data
where the factor $\pi\alpha^2/(2s)$ is represented by a free normalisation
factor N and $\vec{P}_+\vec{P}_-$ is the second free parameter. 
The results of the fit, which was carried out 
over the $\cos\theta_{CM}$ range between $-$0.65 to +0.4, 
\noindent
yielded 
$$\vec{P}_+\vec{P}_-\ =\ 0.044 \ \pm\ 0.126$$
which is consistent within errors with zero thus confirming the identity of
the electron sample as arising predominantly from 
Bhabha scattering. An additional support for the Bhabha scattering
origin of the electrons
is given by
the total number of Monte Carlo electrons emerging
within the $\cos\theta_{CM}$ range of $-$0.65 to +0.4 which is 
consistent
with that predicted by Eq. (\ref{bhapol}) when the low energy
background given in Table \ref{table1} is subtracted. A similar fit
procedure to the Monte Carlo data at 5000 MeV beam energy is
prohibited due to the very large amount of background.

\section{Measurement of the beam polarisation}
For the measurement of the positron longitudinal polarisation we
consider here a single$-$arm polarimeter, with a well defined angular
acceptance, including a pixel detector for the electrons emerging
from the Bhabha process.
The polarisation measurement considered here rests on the two recorded
numbers, 
$N_+$ and $N_-$, which represent respectively the total
Bhabha scattering events detected when the beam$-$target 
polarisation directions are parallel i.e., $\vec{P}_+{\udright}  \vec{P}_-$,
and when they are in opposite directions
i.e., $\vec{P}_+ {\udleft} \vec{P}_-$\ .
By counting $N_+$ and $N_-$ in the $x$ ($\equiv \cos\theta_{CM}$)  range
between $x_{min}$ and $x_{max}$ and using the Bhabha
differential cross section $d\sigma/dx$ defined by Eq. (\ref{maple}),
one obtains an experimental asymmetry
$A_{exp}$ which is given by 
\begin{equation}
A_{exp}=\frac{N_+-N_-}{N_++N_-}=\frac{\int_{x_{min}}^{x_{max}}[d\sigma/dx
(\udright)-d\sigma/dx(\udleft)]dx}
{\int_{x_{min}}^{x^{max}}[d\sigma/dx
(\udright)+d\sigma/dx(\udleft)]dx}=K(x_{max},x_{min})
\vec{P}_+\vec{P}_-\ .
\label{aexp}
\end{equation}
For the range $x_{min}=-0.65$ and $x_{max}=+0.4$ which was chosen here,
the constant $K$ is equal to 0.696.
From this follows that the positron beam 
longitudinal polarisation value is equal to
\begin{equation}
P_+\ =\ \frac{A_{exp}}{KP_-}\ ,
\label{beamp}
\end{equation}
with the two independent error contributions
$$\Delta P_+ = \frac{1}{KP_-}\Delta A_{exp}\ \ \ \rm{and}\ \ \
\Delta P_+ = -\frac{A_{exp}}{KP_-^2}\Delta P_-\ .$$ 
Inasmuch that $P_+\neq 0$ the added in quadrature over all
relative error squared of the measured beam polarisation is 
equal to
\begin{equation}
\left (\frac{\Delta P_+}{P_+}\right )^{2} = \left (\frac{\Delta A_{exp}}{A_{exp}} 
\right )^{2}   + \left (\frac{\Delta P_-}{P_-} \right )^{2} ~, 
\label{error}
\end{equation}
where the statistical error squared of the measured asymmetry is 
\begin{equation}
( \Delta A_{exp})^2 \simeq 4 \frac{N_{+} N_{-}}{N ^3} = 
\frac{1 - A_{exp}^2}{N}=\frac{1}
{{\cal L}T \sigma_x}\left( 1- (KP_+P_-)^2 \right ) ~.     
\label{asy-error}
\end{equation}
Here $\sigma_x$ is the relevant cross section and $N = N_+ + N_-$. 
The luminosity is equal to
${\cal{L}}=d\cdot \rho^{target}_e\cdot N^{beam}_e$ where $d$ is the
thickness
of the target, $\rho^{target}_e$ is the density of electrons in the
target and $N^{beam}_e$ is the number of electrons hitting the target
per second. Finally $T$ is the total measuring time.
For convenience we further consider the case
where the integrated luminosities for the parallel and
anti-parallel polarisations of the positron beam and target electrons
are the same i.e., ${\cal{L}}_+ T_+$=${\cal{L}}_- T_-$.
In this case one has 
\[ N = N_{+} +  N_{-} = {\cal L} \cdot T \cdot \sigma_x  ~,\]
where $\sigma_x$ is the integrated cross section
\[ \sigma_x = \int_{x_{min}}^{x_{max}}\frac{d\sigma}{dx}dx \]\ 

\noindent
which at 250 MeV beam energy is equal to $\sim$2.8 mb for
$x_{min} = -0.65$ and $x_{max} = +0.4$.   
By using  Eqs. (\ref{beamp}) and (\ref{asy-error})
one can rewrite  Eq. (\ref{error}) as follows:
\begin{equation}
 \left (\frac{\Delta P_+}{P_+}\right )^{2}  =
  \frac{1}{{\cal L}T \sigma_x}  \frac{1 - (KP_+P_-)^2}{(KP_+P_-)^2} +
\left (\frac{\Delta P_-}{P_-} \right )^{2}\ . 
\label{plb2_error}
\end{equation}
As long as 
$(KP_+P_- )^2$ \ $\ll\ 1\ $  
one can simplify  Eq. (\ref{plb2_error}) to the form
\begin{equation}
 \left (\frac{\Delta P_+}{P_+}\right )^{2}   \simeq
\frac{1}{{\cal L}T \sigma_x} \frac{1} {(KP_+P_-)^2} 
+ \left (\frac{\Delta P_-}{P_-} \right )^{2}.
\label{plb3_error}
\end{equation}
Finally the time $t_{Int}$ needed to reach a desired relative
beam polarisation precision of $\Delta P_+ /P_+$ is given by 
\begin{equation}
\frac{1}{t_{Int}} \simeq 
{\cal L} \left [ \left ( \frac{\Delta P_+}{P_+}\right )^2 -  \left (\frac{\Delta P_-}{P_-} \right )^{2}\right ]
\left (KP_+P_- \right)^2  \sigma_x ,
\end{equation}
\noindent 
which in turn determines the required number of scattering events,   
$N_{Int} = {\cal L} \times t_{Int} \times \sigma_x$.\\
 
The high precision of $\Delta P_-/P_- =0.5\%$
has already been achieved for the polarisation of the
magnetised iron target in a M\o ller polarimeter which operated
in the JLAB \cite{moller} with an electron beam of 
a few $\mu$A in the energy range of 1 to 6 GeV. 
Thus 
it is expected from Eq. (\ref{plb3_error}) that a low relative statistical
error of\ $\Delta P_+/P_+$
$\simeq$ 0.5$\%$\ 
may be achieved in a short time of a few seconds, 
with a 250 MeV positron beam of 0.1$\mu$A or even less. 
Eventually the over-all precision to be reached by a 
Bhabha polarimeter will thus 
depend mainly on the systematic uncertainties.\\  
 
\section{Summary and conclusions}
The longitudinal polarisation measurement of a positron beam
in a high energy linear collider is required near its production
region to ascertain the polarisation existence and to allow 
the collider team
to optimise its level by providing a fast polarisation measurement
feedback. To satisfy these needs a fixed iron target polarimeter,
the Bhabha polarimeter, is shown to constitute 
an attractive proposition. An iron target as thin as 
10$\mu$m for a fixed target polarimeter has previously been constructed 
and it is expected to reduce to a minimum the various
sources of background to the Bhabha scattering signal.
Inasmuch that the collider design prohibits the presence of
high magnetic fields of
the order of four Tesla which can produce polarisation out-of plane,
the iron polarisation will lie in the plane and 
the target must be tilted. 
Therefore we adopted here the more
realistic case where the iron target is tilted and have shown
that a reliable polarisation measurement can be achieved
even when the target effective
width increases to 30$\mu$m.\\ 

To suppress the major background due to bremsstrahlung events
one should use for the Bhabha scattering identification
and polarisation measurement only the outgoing electrons.
These are sufficient to verify the Bhabha
scattering identity and to supply the data for the polarisation 
measurement of the positron beam. Here one should note that a
similar procedure is not applicable to the M\o ller polarimetry
since both final state charged leptons are electrons.\\

The preferred location of a Bhabha polarimeter in a linear collide
is found to be in
the vicinity where the positron beam reaches the energy of
250 MeV. At this energy the final states are free of 
charged hadrons and at the same time the fraction of 
the non-Bhabha scattering
electrons is still rather small and amounts to some 5.2$\%$ of the
signal even at a target width of 30$\mu$m. 
In addition, at 250 MeV the measured laboratory angular sector
of the outgoing electrons is in the range of  
several degrees and thus, from the engineering side,
is relatively simple to implement. Finally 
the still large Bhabha scattering cross section  
guarantees a low background and a fast measurement feedback for the 
polarisation optimisation effort of the linear collider crew.\\

The option to install a Bhabha polarimeter in the region where
the polarisation beam has an energy in the vicinity of 10 MeV
is prohibited due to the fact that the background merges with the
Bhabha scattering signal events. At 5000 MeV  
the Bhabha polarimetry
is in principle possible however due to the presence of 
hadronic final states and
in particular the relatively large fraction
of the non-Bhabha scattering 
electrons it is a less favourable location for 
a Bhabha polarimetry than that at 250 MeV. 
On the other hand the construction of 
a more elaborated 
polarimeter may allow the removal of the high background at
the 5000 MeV location  
by applying appropriate energy and angle cutoffs. 
The polarisation
measurement duration at 5000 MeV  is expected to be longer 
by a factor of $\sim$20 than that at
250 MeV 
but still short enough to provide a sufficiently fast feedback
to the collider operating team.\\

%\newpage
\noindent
{\large\bf Acknowledgements}\\
\noindent
We would like to thank H. Abramowicz, Y. Ben-Hammou, 
S. Kananov, K. M\"{o}nig, 
S. Riemann, T. Riemann and A. Stahl for many helpful discussions. 
Our thanks are also due to the DESY/Zeuthen particle research
centre and its director U. Gensch, where part of the work reported
here took place.

\end{document}